\documentstyle[aaspp4,emulateapj5,apjfonts]{article}

\gdef\h50min{$h_{50}^{-1}$}
\gdef\kms{km\,s$^{-1}$}

\gdef\3727{[O\,{\sc ii}]\,3727\,\AA}
\gdef\5007{[O\,{\sc iii}]\,5007\,\AA}
\gdef\oii{[O\,{\sc ii}]}
\gdef\oiii{[O\,{\sc iii}]}
\gdef\ms1054{MS\,1054--03}
\gdef\4ang{4000\,\AA}

\gdef\nii{[N\,{\sc ii}]}

\gdef\c695{CDFS-695}
\lefthead{}
\righthead{}
\slugcomment{Accepted for publication in the Astrophysical Journal Letters}
\begin{document}

\title{Gemini Near Infrared Spectrograph Observations of a
Red Star Forming Galaxy at $z=2.225$: Evidence for Shock-Ionization due to
a Galactic Wind\altaffilmark{1}}
\author{
P.~G.~van Dokkum\altaffilmark{2}, 
M.~Kriek\altaffilmark{3},
B.~Rodgers\altaffilmark{4},
M.~Franx\altaffilmark{3},
and
P.~Puxley\altaffilmark{4}
}

\altaffiltext{1}{Based on observations obtained at the Gemini
Observatory, which is operated by the Association of Universities for
Research in Astronomy, Inc., under a cooperative agreement with the
NSF on behalf of the Gemini partnership.}
\altaffiltext{2}{Department of Astronomy, Yale
University, New Haven, CT 06520-8101}
\altaffiltext{3}{Leiden Observatory, P.O. Box 9513, NL-2300 RA, Leiden,
The Netherlands}
\altaffiltext{4}{Gemini Observatory, Casilla 603, La Serena, Chile}

\begin{abstract}

Recent studies
have shown that $K$-luminous galaxies at $2<z<2.5$ have high H$\alpha$
line widths and high \nii\,$\lambda 6583$\,/\,H$\alpha$
ratios. If 
these lines arise from photo-ionization by massive stars
in H\,{\sc ii} regions these results imply that
massive, metal rich galaxies exist at high redshift.
Here we investigate the ionization mechanism in
a galaxy with $K_s=19.1$ and $z=2.225$ in the
Chandra Deep Field South, using the new Gemini
Near Infrared Spectrograph (GNIRS). GNIRS' cross-dispersed mode gives
simultaneous access to
the entire $1\,\mu$m -- $2.5\,\mu$m wavelength range, allowing
accurate measurements of line ratios of distant galaxies. From
the ratio of H$\alpha$\,/\,H$\beta$ we infer that the line emitting
gas is heavily obscured, with $E(B-V)=0.8^{+0.3}_{-0.2}$.
The reddening is
higher than that inferred previously from the UV-optical continuum
emission, consistent with findings for nearby star burst galaxies.
We find that the galaxy has Seyfert-like line ratios,
\nii\,/\,H$\alpha\approx 0.6$ and \oiii\,$\lambda 5007$\,/\,H$\beta \approx 6$, which
can be caused by photo-ionization by
an active galactic nucleus (AGN) or shock ionization due to a strong
galactic wind. Although we cannot exclude the presence of an AGN,
the lack of AGN spectral features in the rest-frame
ultraviolet, the consistency of radio, X-ray, and rest-frame UV
star formation indicators,
the fact that the \oiii\,/\,H$\beta$ ratio
remains high out to $\sim 10$\,kpc from the nucleus,
and the observed gas kinematics all argue for
the wind hypothesis. Similar shock-induced ionization is
seen in nearby star burst galaxies
with strong winds, such as NGC\,1482 and NGC\,3079.
The evidence for shock ionization implies that measurements
of metallicities and dynamical masses of star forming $z>2$ galaxies
should be regarded with caution,
especially since the existence of strong galactic winds
in these objects is well established. Based on Sloan Digital Sky
Survey data for nearby galaxies and the limited data available
at high redshift we speculate that the effects of shocks may
correlate with dust content.
The results presented here demonstrate the importance of measuring
the full rest-frame optical spectra of high redshift galaxies,
and showcase the potential of GNIRS for such studies.

\end{abstract}

\keywords{cosmology: observations ---
galaxies: evolution --- galaxies:
formation
}

\section{Introduction}

Near-infrared (NIR) spectroscopy provides valuable information on the
stellar populations and kinematics of galaxies at $z>2$, as it
gives access to rest-frame optical emission lines that have
been studied extensively in the local Universe. NIR spectra of
optically- and NIR-selected galaxies at $2<z<3$
have constrained or measured their star formation rates, metallicities,
wind velocities, and masses (e.g., {Pettini} {et~al.} 2001; {Erb} {et~al.} 2003; {van Dokkum} {et~al.} 2004; {Shapley} {et~al.} 2004). Among the most surprising results is the finding that
apparently mature
galaxies with H$\alpha$ line widths of $\sigma \gtrsim 200$\,\kms and
\nii\,$\lambda 6583$\,/\,H$\alpha$
ratios of $\gtrsim 0.3$ already existed
at $z\approx 2.5$ ({van Dokkum} {et~al.} 2004; {Shapley} {et~al.} 2004).

The interpretation rests on the assumption that the emission lines
arise from photo-ionization by massive stars in H\,{\sc ii}
regions (see, e.g., {Erb} {et~al.} 2003; {Shapley} {et~al.} 2004). In the local Universe
this assumption is valid for the majority of star forming
galaxies, but not for all. The \nii\,/\,H$\alpha$ ratio
can be influenced by photo-ionization by an AGN continuum,
shock ionization caused by the interaction of outflowing gas with the
ambient interstellar medium,
photo-ionization by hot, high metallicity stars in a dense environment,
and other processes (see,
e.g., {Veilleux} {et~al.} 1994, and references therein). The potential effects of
shock ionization are particularly relevant, as the existence of
large scale outflows in star forming galaxies at $z>2$
is well established (e.g., {Franx} {et~al.} 1997; {Pettini} {et~al.} 1998, 2001).
The ionization source cannot be constrained from \nii\,/\,H$\alpha$
alone (although
values larger than $\approx 0.5$ point to other mechanisms
than photo-ionization in H\,{\sc ii} regions), and the ``standard''
method of spectral
classification relies on combinations of line ratios such
as \nii\,/\,H$\alpha$ and \oiii\,$\lambda 5007$\,/\,H$\beta$
(e.g., {Baldwin}, {Phillips}, \&  {Terlevich} 1981; {Veilleux} \& {Osterbrock} 1987).

Measuring
these line combinations at high redshift is difficult due to
the limited wavelength
coverage of most existing
NIR spectrographs on large telescopes.
In this {\em Letter}, we use the newly
commissioned Gemini Near Infrared Spectrograph (GNIRS) to study
the line ratios of a $K$-selected star forming galaxy at
$z=2.225$ in the Chandra Deep Field South (CDF-S).
GNIRS can be operated in cross-dispersed mode, which produces
spectra over
the entire 1\,$\mu$m\,--\,2.5\,$\mu$m wavelength range at a resolution
of $R \sim 1800$.
GNIRS thus allows simultaneous measurement of rest-frame
optical lines falling in the $J$, $H$, and $K$ bands through identical
atmosphere and aperture on the sky, and hence accurate and
efficient determinations of line ratios.

\begin{figure*}[t]
\epsfxsize=14.6cm
\epsffile[-90 135 566 658]{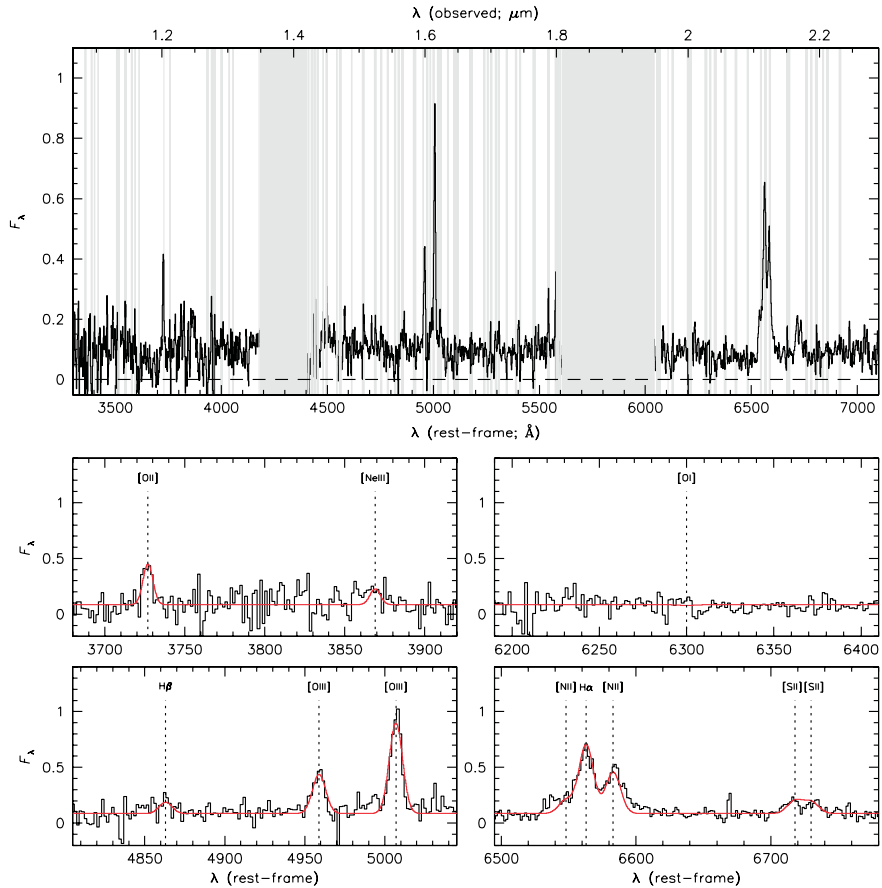}
\caption{\small
Cross-dispersed low resolution spectrum of \c695\ at $z=2.225$
(top panel), smoothed
with a 200\,\kms\ boxcar filter. Grey bands indicate regions of
strong atmospheric emission or
absorption. The familiar rest-frame optical
emission lines are readily identified, showing the potential of
GNIRS for measuring redshifts and line ratios in the near-IR.
The bottom panels show (unsmoothed) details of the spectrum,
centered on emission lines.
\label{spec.plot}}
\end{figure*}

\section{Spectroscopy}

We observed galaxy \c695\ on 2004 September 2--3 UT with GNIRS on
Gemini-South, under mostly clear skies with image quality
$\approx 0\farcs 7$.
The galaxy was selected
because of its $K$-magnitude and red $J-K_s$ color. It has $K_s=19.1$ and
$J-K_s=2.3$ (S.\ Wuyts et al., in preparation), and
is one of the most luminous members of the Distant Red
Galaxy (DRG) class ({Franx} {et~al.} 2003;
{van Dokkum} {et~al.} 2003) in the CDF-S.
After measuring the redshift with GNIRS we realized that
the galaxy is object 5 in the {Daddi} {et~al.} (2004) sample of
$K$-selected star forming galaxies.  We note that {Daddi} {et~al.} (2004)
give a slightly bluer color of $J-K_s=2.15$ for this object; we
cannot determine the source of this small discrepancy.

GNIRS was operated in cross-dispersed mode, using the short camera
with the 32\,lines\,mm$^{-1}$ grating and the $0\farcs 68$ slit.
Between individual $\sim 300$\,s exposures
the objects was
moved along the $6\farcs 2$ slit in an ABA$'$B$'$ pattern, with A$'$
and B$'$ small ($0\farcs 3$) perturbations of A and B.
The total exposure time was 5500\,s.
The data reduction used a combination of standard procedures
and custom software, and is described in M.\ Kriek et al., in preparation.
The reduced spectrum
is shown in Fig.~\ref{spec.plot}. 
The continuum is clearly detected along with familiar rest-frame
optical emission lines, which provide a redshift $z=2.225$.
This high quality spectrum demonstrates the potential of
GNIRS for studies of faint objects, and specifically for
measuring redshifts of $K$-selected galaxies.

Line fluxes were determined from the residual emission after
subtraction of a linear fit to the continuum, with errors
estimated from repeat
measurements in empty regions. Equivalent widths were
determined from a direct comparison of the line fluxes to the
continuum fits, and the line fluxes were calibrated by
comparing the measured total $K_s$ magnitude to a synthetic magnitude
determined from the GNIRS spectrum. This method assumes
that the line emitting gas has the same distribution as the continuum
emission, and we estimate that the systematic error is
$\approx 20$\,\%. Finally, line luminosities were calculated
using $\Omega_m=0.3$, $\Omega_{\Lambda}=0.7$, and
$H_0=70$\,\kms\,Mpc$^{-1}$. Fluxes, luminosities, and rest-frame
equivalent widths are listed in Table~1. Errors do not
include a $\sim 20$\,\% uncertainty
in the overall normalization of the spectrum (which cancels
for line ratios).

\section{Reddening}

The reddening of the line-emitting gas can be derived from the
Balmer decrement, after correcting for underlying continuum
absorption at H$\alpha$ and H$\beta$. Stellar populations with
ages $0.5-2$\,Gyr (F\"orster Schreiber et al.\ 2004;
Daddi et al.\ 2004) have Balmer absorption equivalent widths of
$4\pm 1$\,\AA. The corresponding
corrections to the measured
line luminosities are a factor of $1.04$ for
H$\alpha$ and a factor of $1.3$ for H$\beta$, and
the corrected ratio H$\alpha$\,/\,H$\beta = 6.6^{+2.6}_{-1.6}$. Assuming
an intrinsic ratio of 2.76, which is valid for H\,{\sc ii}
regions and also holds to good approximation for shock
ionized regions ({Binette}, {Dopita}, \& {Tuohy} 1985), the reddening
$E(B-V) = 0.8^{+0.3}_{-0.2}$.
This large value is consistent with previous indirect
determinations for DRGs (van Dokkum et al.\ 2004;
F\"orster Schreiber et al.\ 2004).

The reddening of the
line-emitting gas can be compared to that of the stellar continuum
emission. {Daddi} {et~al.} (2004) derive
$E(B-V)=0.4$ for the continuum of \c695, from
SED fits. This difference between the reddening of the line-emitting
gas and the continuum is
consistent with previous results for
nearby star burst galaxies ({Calzetti} 1997) and DRGs
({van Dokkum} {et~al.} 2004).

\vbox{
\begin{center}
\leavevmode
\hbox{%
\epsfxsize=8cm
\epsffile{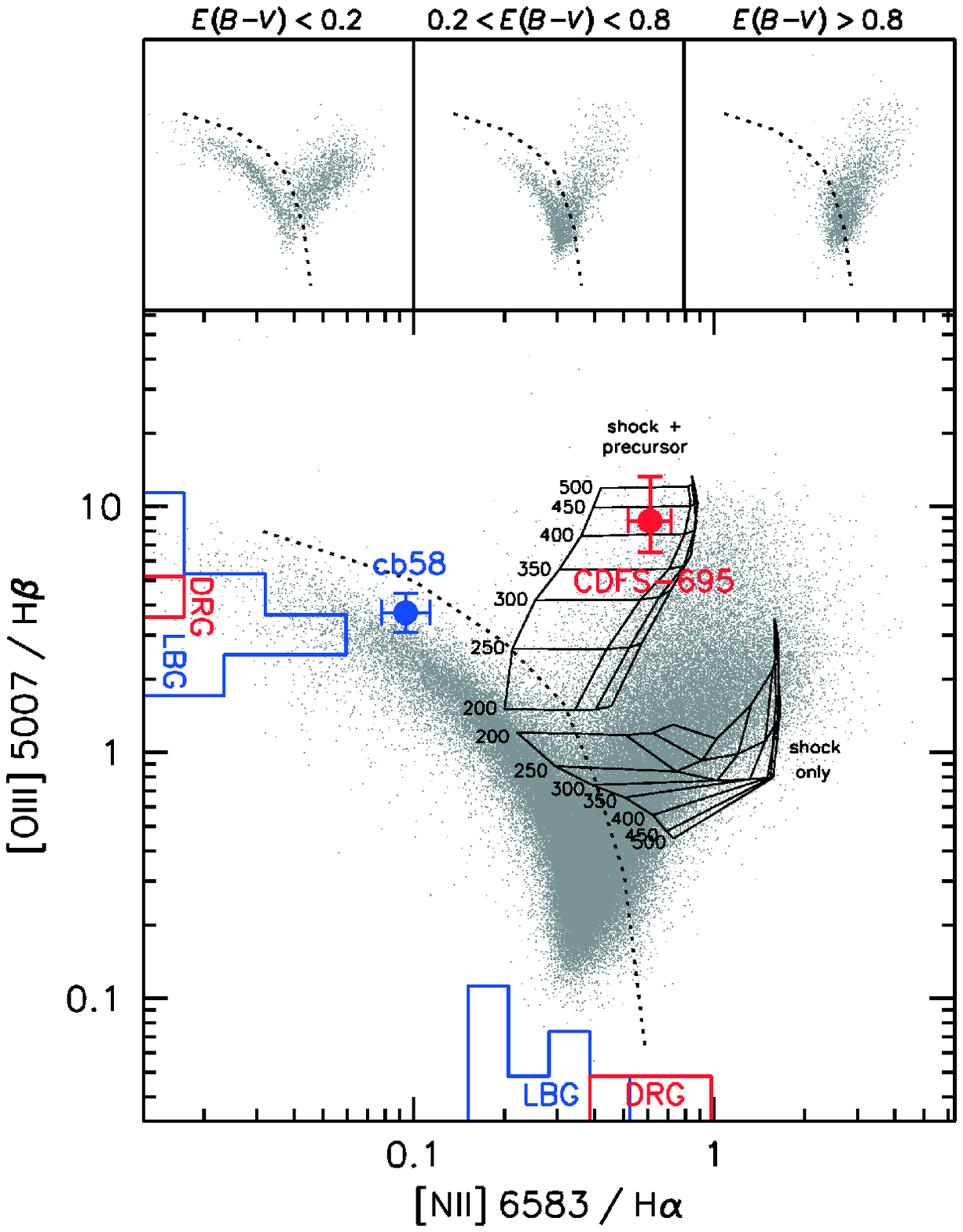}}
\figcaption{\small
Diagnostic diagram for spectral classification. Small dots are
nearby galaxies from the SDSS. Galaxies to the left of the dotted
line are normal star forming galaxies (Kauffmann et al.\ 2003).
The red symbol is \c695, which has a Seyfert-like spectrum. 
The grids
show shock ionization models from Dopita
\& Sutherland (1995), with and without a precursor region.
Histograms show measurements for DRGs (van Dokkum et al.\ 2004)
and LBGs at $z\sim 2$ (\nii\,/\,H$\alpha$;
Shapley et al.\ 2004)
and $z\sim 3$ (\oiii\,/\,H$\beta$; Pettini et al.\ 1998; 2001).
The blue symbol is the LBG MS\,1512-cB58 (Teplitz et al.\ 2000).
The top panels show SDSS galaxies binned by reddening: the
most dusty galaxies often have LINER/Seyfert-like line ratios.
\label{lineratios.plot}}
\end{center}}

\section{Spectral Classification}

Galaxies whose emission lines arise from photo-ionization in H\,{\sc
ii} regions follow a well-defined sequence in diagrams involving the
line ratios \oiii\,/\,H$\beta$, \nii\,/\,H$\alpha$, [S\,{\sc
ii}]\,/\,H$\alpha$, and [O\,{\sc i}]\,/\,H$\alpha$, largely
because of
strong correlations between ionization parameter and metallicity
(e.g., {Veilleux} \& {Osterbrock} 1987).  The
diagnostic diagram of \oiii\,$\lambda
5007$\,/\,H$\beta$ versus \nii\,$\lambda 6583$\,/\,H$\alpha$ is
particularly useful, as these lines are relatively strong and
reddening corrections are small.

Figure \ref{lineratios.plot} shows
the distribution of nearby galaxies in this diagram using data
from the Sloan Digital Sky Survey DR2 (Kauffmann et al.\ 2003;
Tremonti et al.\ 2004). The line ratios were
corrected for reddening, and only the $\sim 124,000$
galaxies for which all four lines
are detected at $>2 \sigma$ are shown.
The dotted line shows the separation
between H\,{\sc ii} type emission and LINER/Seyfert type emission
as adopted by {Kauffmann} {et~al.} (2003). The emission lines of
galaxies to the right of this line are affected or dominated
by other sources of ionization than massive stars -- in most
cases an active nucleus.

The solid red symbol in Fig.\ \ref{lineratios.plot}
shows the reddening-corrected lineratios of \c695. Its
\nii\,/\,H$\alpha$ ratio of $\sim 0.6$ and
\oiii\,/\,H$\beta$ ratio of $\sim 6$ place the galaxy
firmly in the Seyfert region (see, e.g., {Veilleux} \& {Osterbrock} 1987).

\begin{small}
\begin{center}
{ {\sc TABLE 1} \\
\sc Emission Lines} \\
\vspace{0.1cm}
\begin{tabular}{lccccc}
\hline
\hline
Line & $\lambda_{\rm rest}$ & $W_{\lambda}$\,(\AA) & $F^a$ & $L^b$ &
$\Delta^c$ \\
\hline
     \oii\  &  3727  & 33 &   8.0 &    3.0 &      0.2 \\
$[$Ne\,{\sc iii}$]$\     &   3869 & 13 &   3.2 &   1.2 &      0.4 \\
      H$\beta$ &   4863  &  12 &  2.9  &        1.1 &       0.4 \\
      \oiii\ &        4959 &  43 &  10  &     3.9 &     0.1 \\
      \oiii\ &        5007  &     99 &  24 &       9.0 &     0.1\\
$[$O\,{\sc i}$]$ & 6300 & $<8$ & $<2$ & $<0.7$ & --- \\
      \nii\ &        6548  &     27 &    6.4 &       2.4 &      0.2 \\
     H$\alpha$ &     6563    & 99 & 24 & 9.0 & 0.1\\
       \nii\ &       6583   &    60 &   14 &       5.4 &     0.1\\
       $[$S\,{\sc ii}$]$ &  6718   &    19 &  4.5 &       1.7 &      0.2\\
       $[$S\,{\sc ii}$]$ & 6730   &     17 &  4.1  &      1.6 &      0.2 \\
\hline
\end{tabular}
\end{center}
$^a$\,Flux in units of $10^{-17}$\,ergs\,s$^{-1}$\,cm$^{-2}$.\\
$^b$\,Luminosity in units of $10^{42}$\,ergs\,s$^{-1}$.\\
$^c$\,Relative uncertainty in flux and luminosity.
\end{small}

\section{Ionization Mechanism}
\label{kin.sec}

The most straightforward interpretation is that the line emission
arises from ionization by the power-law continuum of an AGN.
As the line widths are only $\approx 250$\,\kms\ the galaxy would
be a Type II quasar (QSO) (e.g., {Stern} {et~al.} 2002), with the broad line
region hidden by dust. However, other explanations are also possible.
In particular, ionization due to fast radiative shocks naturally
produces LINER or Seyfert-like line ratios
(see, e.g., {Dopita} \& {Sutherland} 1995).

Whereas it is difficult to unambiguously identify a Type II
AGN in the rest-frame optical, it can be straightforward in X-rays,
radio, and/or the rest-frame ultra-violet (UV). The rest-UV spectrum
is especially important: Seyfert II galaxies (and known
type II QSOs) typically have
featureless continua with strong emission lines of Ly$\alpha$ and
high-ionization species such as N\,{\sc v}, C\,{\sc iv}, and [Ne\,{\sc
iv}] ({Kinney} {et~al.} 1991).  {Daddi} {et~al.} (2004) obtained a high quality
rest-frame UV spectrum of \c695. The spectrum is similar to those
of normal star forming galaxies, and shows no AGN features:
Ly$\alpha$ is in absorption, and familiar rest-frame UV absorption
lines such as Si\,{\sc iv}\,$\lambda 1393, 1402$ are strong.
The star formation rate implied by the de-reddened UV continuum is
$\sim 500 M_{\odot}$\,yr$^{-1}$, in good agreement with the
star formation rate implied by its (soft)
X-ray flux and its radio emission ({Daddi} {et~al.} 2004). In summary, there
is no evidence for the presence of an
AGN other than the unusual rest-frame optical
line ratios.

Although we cannot exclude a hidden AGN (or an AGN that has recently
turned off) as the ionization source,
shock ionization is a plausible alternative.
The solid line in Fig.\ \ref{lineratios.plot} shows
shock models from {Dopita} \& {Sutherland} (1995). The ``shock
+ precursor'' models predict similar line ratios as those
observed. In these models
the high \nii\,/\,H$\alpha$ ratios are produced by the shock itself,
and the high \oiii\,/\,H$\beta$ ratios arise from
photo-ionization of preshock regions by extreme UV and X-ray photons generated
in the shock.

The shock interpretation is supported by the behavior of the emission
lines at large distances from the nucleus. As shown in Fig.\
\ref{2dspec.plot} the lines can be traced to $\sim 10$\,kpc from
the nucleus, and the line ratios remain Seyfert-like: we find
\nii\,/\,H$\alpha \approx 0.3$ and \oiii\,/\,H$\beta \approx 10$
in the outer regions of the galaxy. The null or inverted
gradient in \oiii\,/\,H$\beta$ is difficult
to explain if photo-ionization from the nucleus is the cause.

Furthermore, there is direct evidence for a strong galactic
wind which can provide the mechanical energy for the shocks:
the velocity difference between rest-frame
UV absorption lines (Daddi et al.\ 2004)
and rest-frame optical emission lines implies a wind velocity
of $\sim 300$\,\kms, and the width and tilt of the
rest-frame optical emission lines
demonstrate gas motions of $\gtrsim 300$\,\kms.
A natural origin of the wind is the combined effects of
supernovae and winds from massive stars.
Local analogs to \c695\ could be NGC\,3079, which has a windblown
``superbubble'' of diameter 1.1\,kpc displaying similar
line ratios as \c695\ ({Veilleux} {et~al.} 1994),
and NGC 1482, in which the entire (bi-polar) wind appears
to be shock-excited (Veilleux \& Rupke 2002).

\vbox{
\begin{center}
\leavevmode
\hbox{%
\epsfxsize=7cm
\epsffile{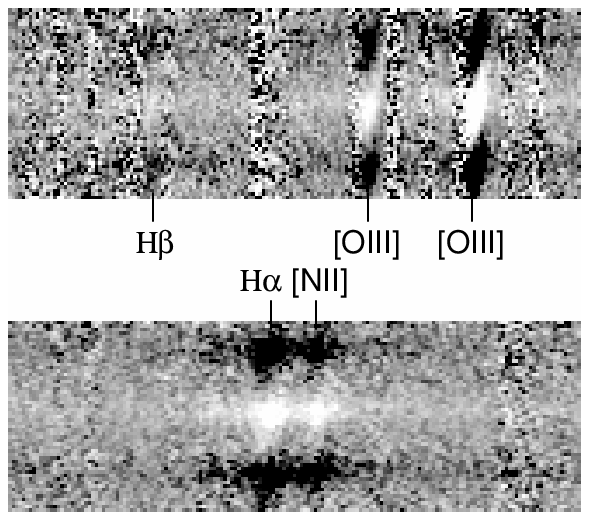}}
\figcaption{\small
Two-dimensional spectrum showing position along the $y$-axis
and wavelength along the $x$-axis, in the vicinity of
$1.6\,\mu$m and
$2.1\,\mu$m. Negative images are caused by the dither pattern
within the $6''$ slit. The emission lines are clearly extended in
the spatial direction, and show a pronounced velocity gradient.
Line ratios remain Seyfert-like
even at large distances from the center of the galaxy.
\label{2dspec.plot}}
\end{center}}

\section{Discussion}

Whatever the cause of the Seyfert-like line ratios, the implication is
that the interpretation of the rest-frame optical emission lines in
this galaxy is more complicated than is usually assumed.  Metallicity
indicators such as \nii\,/\,H$\alpha$, the $R_{23}$ index, and the
\oiii\,/\,\nii\ index (e.g., Pettini \& Pagel 2004) would produce
erroneous results. Using only the $H$-band spectrum we
would derive a low metallicity, and using only the $K$-band spectrum
we would infer a high metallicity.
The H$\alpha$ kinematics are probably also influenced by
the outflow and may not trace the potential
({Veilleux} {et~al.} 1994). Finally, the H$\alpha$ luminosity can be
shock-enhanced
and cannot simply be converted to a star formation rate.
The extinction-corrected H$\alpha$ luminosity, $L_{{\rm H}\alpha}
\sim 1.0\times 10^{44}$\,ergs\,s$^{-1}$,
would imply a star formation rate of
$\sim 800\,M_{\odot}$\,yr$^{-1}$ (Kennicutt 1998). However,
the luminosity produced in the shock is of
the same order as the observed luminosity:
$L_{{\rm H}\alpha} \sim
2.4 \times 10^{26} n_0 A V_s^3$, with
$n_0 \sim 10^2$\,cm$^{-3}$, $V_s \sim 300$\,\kms, and
$A \sim 10^8$\,pc$^2$
(Binette et al.\ 1985).

A critical question is whether \c695\ is exceptional or typical.
Its \nii\,/\,H$\alpha$ ratio is similar to the three other DRGs
for which these lines have been measured
(see Fig.\ \ref{lineratios.plot}), suggesting LINER/Seyfert
like line ratios may be common in red $z>2$ galaxies.
In Lyman break galaxies (LBGs) at $z\sim 2$ this ratio is
usually lower (Erb et al.\ 2003),
even for the subsample with $K<20$ ({Shapley} {et~al.} 2004).
It is interesting
to note that the \oiii\,/\,H$\beta$ ratios of LBGs at $z\approx 3$
are very high (Pettini et al.\ 2001; see Fig.\ \ref{lineratios.plot}),
although still consistent
with the low metallicity tail of the distribution of normal
galaxies. The only LBG for which both line ratios have been measured so
far is the lensed $z=2.7$ galaxy cb58 (Teplitz et al.\ 2000).
It falls on the relation defined by normal star forming galaxies
in Fig.\ \ref{lineratios.plot}, but there is a factor of
$\sim 5$ difference
between the N abundance inferred from the \nii\ emission and from
the N\,{\sc i} absorption in the rest-frame UV (Pettini
et al.\ 2002). This may hint at low-level effects of shocks,
but other explanations are also possible.

The effects of shocks and possibly AGN may be more important
for the dusty DRGs than for LBGs.
The top panels in Fig.\
\ref{lineratios.plot} show the distributions of SDSS galaxies
binned by reddening. Interestingly,
a large fraction of the most obscured galaxies show LINER and
Seyfert-like line ratios, qualitatively consistent with earlier results
for luminous infrared galaxies (Heckman et al.\ 1987).
Submm galaxies at high redshift exhibit a large range in
\nii\,/\,H$\alpha$ ratio (Swinbank et al.\ 2005), and it will be
interesting to see where they fall on diagnostic diagrams like
Fig.\ \ref{lineratios.plot}. More generally,
it is clearly important to measure combinations of line
ratios for a large, homogeneous sample of $z\sim 2.5$ galaxies, and to
investigate correlations with dust content,  kinematics, stellar mass
and nuclear activity.
As demonstrated in this {\em Letter}, Gemini/GNIRS is very well
suited for such studies.

\acknowledgements{We thank Max Pettini for insightful comments
on the manuscript.
}

\end{document}